\documentclass{iaus}
\usepackage{graphicx, natbib}

\newcommand{\aap}   {\textit{A\&A}}
\newcommand{\aaps}  {\textit{A\&AS}}
\newcommand{\aj}    {\textit{AJ}}
\newcommand{\apj}   {\textit{ApJ}}
\newcommand{\apjl}  {\textit{ApJ}}
\newcommand{\apjs}  {\textit{ApJS}}
\newcommand{\araa}  {\textit{ARA\&A}}
\newcommand{\mnras} {\textit{MNRAS}}

\title[The Nature of Stellar Winds] 
{The Nature of Stellar Winds in the Star-Disk Interaction}

\author[Matt \& Pudritz]   
{Sean Matt$^1$ \and Ralph E. Pudritz$^2$}

\affiliation{$^1$Dept.\ of Astronomy, U.\ of Virginia
PO Box 400325, Charlottesville, VA 22904-4325, USA \break 
email: spm5x@virginia.edu \\[\affilskip]
$^2$Physics and Astronomy Dept., McMaster University, 
Hamilton, ON, L8S 4M1, Canada \break 
email: pudritz@physics.mcmaster.ca}

\pubyear{2007}
\volume{243}  
\pagerange{1--8}
\date{June 30, 2007}
\jname{Star-Disk Interaction in Young Stars}
\editors{A.C. Editor, B.D. Editor \& C.E. Editor, eds.}
\begin{document}

\maketitle

\begin{abstract}
Stellar winds may be important for angular momentum transport from
accreting T Tauri stars, but the nature of these winds is still not
well-constrained.  We present some simulation results for
hypothetical, hot ($\sim 10^{6}$ K) coronal winds from T Tauri stars,
and we calculate the expected emission properties.  For the high mass
loss rates required to solve the angular momentum problem, we find
that the radiative losses will be much greater than can be powered by
the accretion process.  We place an upper limit to the mass loss rate
from accretion-powered coronal winds of $\sim 10^{-11} M_\odot$
yr$^{-1}$.  We conclude that accretion powered stellar winds are still
a promising scenario for solving the stellar angular momentum problem,
but the winds must be cool (e.g., $~10^{4}$ K) and thus are not driven
by thermal pressure.

\end{abstract}

\firstsection 

\section{Introduction} \label{sec_intro}

Observations \citep[e.g.,][]{herbstea07} reveal that a large fraction
of accreting T Tauri stars (CTTSs) spin slowly, that is at $\sim 10$\%
of breakup speed.  This is surprising because the accretion of disk
material adds angular momentum to the star
\citep[e.g.,][]{mattpudritz07a}.  One promising scenario to explain
how the slowly spinning stars rid themselves of this accreted angular
momentum, proposed by \citet{hartmannstauffer89}, is that a stellar
wind carries it off.  For this to work, the mass outflow rate should
be approximately proportional to the accretion rate.  Depending on the
stellar magnetic field strength (among other things), in order to
solve the stellar angular momentum problem, the wind outflow rate
needs to be of the order of 10\% of the accretion rate
\citep{mattpudritz05}.

Since the ``typical'' mass accretion rate observed in the CTTSs is
$\dot M_{\rm a} \sim 10^{-8} M_\odot$ yr$^{-1}$
\citep{johnskrullgafford02}, this means the stellar wind should have a
mass outflow rate of $\dot M_{\rm w} \sim 10^{-9} M_\odot$ yr$^{-1}$.
A wind this massive requires a lot of power to accelerate it, and
\citet{mattpudritz05} suggested that a fraction of the potential
energy liberated by the accretion process goes into driving the wind.
In the case of a coronal wind (i.e., $\sim 10^6$ K, thermally driven),
for example, this requires $\sim 10$\% of the accretion power
\citep{mattpudritz05}.  There is some observational evidence for
accretion-powered stellar winds in these systems \citep{edwardsea06,
kwan3ea07}.

But what is the nature of T Tauri stellar winds?  How massive are
they, and what drives them?  The mass outflow rates of stellar winds
is very poorly constrained observationally
\citep[e.g.,][]{dupreeea05}.  This is basically due to the extreme
difficulty in disentangling the signatures of a stellar wind from
signatures of a wind from the inner edge of a disk and a host of other
energetic phenomena exhibited by CTTSs.  The wind driving mechanism is
also not constrained and is the primary focus of this paper.

\section{The T Tauri Coronal Wind Hypothesis} \label{sec_hypothesis}

T Tauri stars are magnetically active and possess hot, energetic
corona \citep[for a review, see][]{feigelsonmontmerle99}.  They are
4--5 orders of magnitude more luminous in X-rays than the sun.  Thus,
it stands to reason that they drive solar-like coronal winds, but more
powerful.  In this case, the wind is primarily thermal
pressure-driven, and the wind temperature needs to be $\sim 10^6$ K
for the pressure force to overcome gravity.  As a first step, we make
the hypothesis here that some of the accretion power is transferred to
heat in the stellar corona, and thus drives a coronal wind.

There is only one calculation in the literature (that we are aware of)
that constrains the mass outflow rate of coronal winds from CTTSs.
Specifically, \citet{bisnovatyikoganlamzin77} calculated the X-ray
emission from coronal winds.  From these calculations,
\citet{decampli81} concluded that, in order for the wind emission to
be consistent with the observed X-ray luminosities, the outflow rate
of a T Tauri coronal wind must be less than $\sim 10^{-9} M_\odot$
yr$^{-1}$.  As discussed above, a wind this massive may still be
important for angular momentum transport, and thus we proceed.

\section{Coronal Wind Simulations} \label{sec_simulations}

To calculate realistic wind solutions, we carried out 2.5D
(axisymmetric) ideal magnetohydrodynamic (MHD) simulations of coronal
winds.  For simplicity, we did not include the accretion disk.  We
employ the numerical code and method described by
\citet{mattbalick04}.  This allows us to obtain steady-state wind
solutions for a Parker-like coronal wind \citep{parker58}, as modified
by the presence of stellar rotation and a rotation-axis-aligned dipole
magnetic field.  We assume a polytropic equation of state ($P \propto
\rho^\gamma$), with no radiative cooling.  The fiducial parameters are
given in table \ref{tab_parms}, adopted to represent values for a
``typical'' CTTS.

\begin{table}[h]
  \begin{center}
\caption{Fiducial Stellar Wind Parameters \label{tab_parms}}

\begin{tabular}{ll}\hline
Parameter                 & Value \\ \hline
$M_*$                     & 0.5 $M_\odot$ \\
$R_*$                     & 2.0 $R_\odot$ \\
$B_*$ (dipole)            & 200 G         \\
$f$                       & 0.1           \\
$\dot M_{\rm w}$ & $1.9 \times 10^{-9} M_\odot$ yr$^{-1}$ \\
$T_{\rm c}$               & $1.3 \times 10^6$ K          \\
$\gamma$                  & 1.40           \\ \hline
\end{tabular}

  \end{center}
\end{table}

In table \ref{tab_parms}, $M_*$ and $R_*$ are the stellar mass and
radius; $B_*$ is the magnetic field strength of the dipole magnetic
field at the surface and equator of the star; $f$ is the spin rate of
the star, expressed as a fraction of the breakup rate; $T_{\rm c}$ is
the temperature at the base of the corona; and $\gamma$ is the
polytropic index.

\begin{figure}
 \begin{center}
 \includegraphics{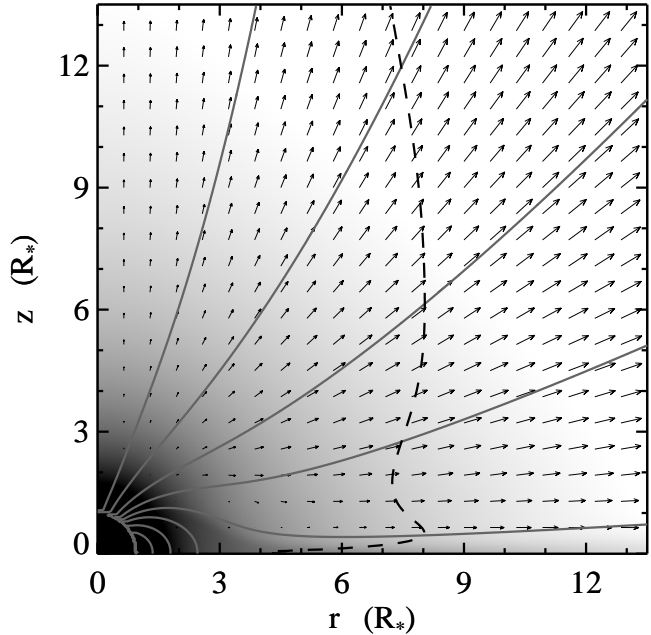}
  \caption{Greyscale of log density, velocity vectors, and magnetic
  field lines illustrate the structure of the steady-state wind
  solution for our fiducial case.  The dashed line represents the
  Alfv\'en surface, where the wind speed equals the local Alfv\'en
  speed.  The rotation axis is vertical, along the left side of the
  plot.}\label{fig_simulation}
 \end{center}
\end{figure}

Figure \ref{fig_simulation} illustrates the steady-state wind solution
for the fiducial case.  We find that this wind carries away enough
angular momentum to counteract the spin up torque from an accretion
rate of $\dot M_{\rm a} \approx 5 \times 10^{-9} M_\odot$ yr$^{-1}$.
We also carried out a parameter study (Matt \& Pudritz 2007, in
preparation), which generally validates the idea that a stellar wind
can indeed remove the accreted angular momentum in CTTSs.

\section{Emission Properties of Coronal Wind} \label{sec_emission}

The simulation results of the previous section provide detailed
solutions for the density and temperature in coronal stellar winds.
Although the simulations did not include radiative cooling effects, it
is instructive to examine the emission properties expected from these
winds, ex post facto.  For this, we employ the CHIANTI line database
and IDL software \citep{dereea97, landiea06}, which allows us to
calculate spectra and total radiative cooling rates in the wind.

The CHIANTI package assumes, among other things, that the ionization
and excitation levels in the plasma are in a steady-state; all lines
are optically thin; the plasma is in coronal equilibrium, so that the
ionization state is in LTE.  These assumptions are appropriate for the
purposes of this work, and we also adopt cosmic abundances for the
gas.

     \subsection{Illustrative Synthetic Spectrum}

\begin{figure}
 \includegraphics[angle=90,scale=.57]{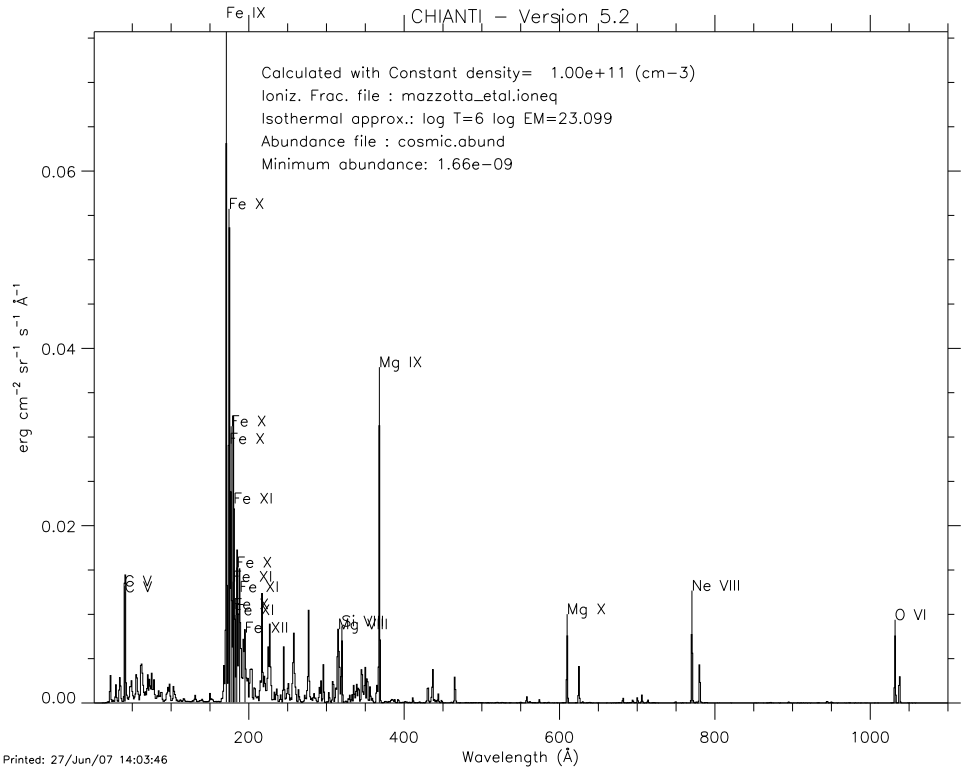}
 \caption{Synthetic spectrum of isothermal, $10^6$ K, optically thin
 coronal plasma.  Flux units are arbitrary.  The figure is generated
 by CHIANTI software.}\label{fig_spectrum}
\end{figure}

For illustrative purposes, figure \ref{fig_spectrum} shows a spectrum,
computed by CHIANTI, of an isothermal plasma with a temperature of
$10^6$ K.  It is clear that the cooling is dominated by line emission.
In this case, the three strongest emission lines (of Fe IX 171.1 \AA,
Fe X 174.5 \AA, and Mg IX 368.1 \AA) account for approximately $20$\%
of the total luminosity.  Furthermore, only about 1\% of the total
energy is emitted shortward of 30 angstroms (i.e., in X-rays), and the
vast majority of the emission is in the extreme UV.

     \subsection{Total Radiative Losses}


CHIANTI also provides a tool to calculate the total cooling rate
(i.e., radiated luminosity in erg s$^{-1}$; which is essentially an
integration of the emission spectrum over wavelength, etc.) for any
given coronal density, temperature, and emitting volume.  With this,
we calculate the cooling in each computational gridcell of our
simulations, and sum over all gridcells, to obtain the total
luminosity of the simulated wind solution.  For the fiducial case, the
total wind luminosity is a few times $10^{34}$ erg s$^{-1}$.  Since
optically thin emission is proportional to density squared, and since
the mass outflow rate in the wind is approximately proportional to
density, we express the luminosity of the wind as
\begin{eqnarray}
  \label{eqn_lw}
L_{\rm w} \sim 10^{34} \;\; {\rm erg \; s}^{-1} \;\;
               \left({\dot M_{\rm w} \over 
                     10^{-9} M_\odot {\rm \; yr}^{-1}}\right)^2.
\end{eqnarray}

As suggested by the example spectrum (fig.\ \ref{fig_spectrum}), if
$\sim 1$\% of this emission is emitted in X-rays, the X-ray luminosity
of the wind is $\sim 10^{32}$ erg s$^{-1}$.  This is significantly
higher than the typically observed X-ray luminosity of CTTSs of $\sim
10^{30}$ erg s$^{-1}$ \citep{feigelsonmontmerle99}.  Of course, we
have calculated the total cooling rate, which is not exactly the
observed luminosity.  Consider that approximately half of this
radiation will be blocked by the star, and there will likely be
significant absorption of these soft X-rays.  Still, it does not seem
avoidable that the predicted X-ray luminosity from the fiducial
coronal wind solution is much higher than typically observed.

     \subsection{Accretion Power}

More importantly, we must consider the energy budget of the wind.  The
total cooling rate, $L_{\rm w}$, of the fiducial wind is two orders of
magnitude larger than the kinetic energy in the wind ($0.5 \dot M_{\rm
w} v_\infty^2$, where $v_\infty^2$ is the wind speed)---this is
approximately equivalent to saying that the cooling time is two orders
of magnitude shorter than the wind acceleration time.  Thus, it takes
a lot more energy to keep this plasma hot (while it radiates) than it
does to accelerate the matter away from the star.

In the accretion-powered stellar wind scenario, the energy in the wind
somehow derives from the gravitational potential energy released by
accreting gas ($\sim G M_* \dot M_{\rm a} / R_*$).  This accretion
power, assuming the fiducial stellar mass and radius, can be expressed
approximately as
\begin{eqnarray}
  \label{eqn_la}
L_{\rm a} \sim 10^{32} \;\; {\rm erg \; s}^{-1} \;\;
               \left({\dot M_{\rm a} \over 
                     10^{-8} M_\odot {\rm \; yr}^{-1}}\right).
\end{eqnarray}
As discussed in section \ref{sec_intro}, in order for stellar winds to
solve the angular momentum problem, torque balance requires $\dot
M_{\rm w} / \dot M_{\rm a} \sim 0.1$.  Thus, if we fix this ratio of
mass flow rates, it is clear from equations \ref{eqn_lw} and
\ref{eqn_la} that there is not enough accretion energy to keep coronal
winds hot, in the fiducial case.

     \subsection{An Upper Limit on T Tauri Coronal Winds}

If we fix the ratio $\dot M_{\rm w} / \dot M_{\rm a} \sim 0.1$, it is
evident from equations \ref{eqn_lw} and \ref{eqn_la} that there will
be enough accretion power to drive a coronal wind when the wind
outflow rate is
\begin{eqnarray}
  \label{eqn_mdotlimit}
\dot M_{\rm w} \;\; {}^<_\sim \;\; 10^{-11} \;\; M_\odot \; {\rm yr}^{-1}.
\end{eqnarray}
Thus, in principle, accretion-powered coronal winds can remove the
accreted angular momentum for $\dot M_{\rm a} \sim 10^{-10} M_\odot$
yr$^{-1}$.  However, for accretion rates this low, the spin up torque
from accretion is so small that the time for the star to spin up from
this torque is comparable to the entire pre-main-sequence lifetime
\citep[e.g.,][]{mattpudritz07a}.  In other words, for these low
accretion rates, there is no angular momentum problem.  The logical
conclusion is that, in order for accretion-powered stellar winds to
solve the angular momentum problem, the winds cannot be as hot as we
have considered here.

     \section{On the Validity of Our Simulated Wind Solutions}

We showed in section \ref{sec_emission} that the expected emission
properties of our fiducial, coronal winds effectively rules them out.
In other words, our assumption in this paper that the wind is driven
by thermal-pressure is not realistic.  However, it is important to
note that the angular momentum carried in the wind does not depend on
what drives the wind.  Instead, the angular momentum outflow rate
depends only on $B_*$, the rotation rate, $\dot M_{\rm w}$, $R_*$, and
the wind velocity.  As long as ``something'' accelerates the wind to
speeds similar to what we see in our simulations (of the order of the
escape speed), the torque we calculate is approximately correct.

For example, if the wind is cold and driven by Alfv\'en waves, the
driving force can be parameterized as being proportional to $-\nabla
\xi$ \citep[where $\xi$ is the wave energy density;][]{decampli81}.
This has the same functional form as the thermal-pressure force ($-
\nabla P$) used in our simulations, so there is some form of $\xi$
that will result in a wind solution with exactly the same density and
kinematics as our simulations (but a different temperature).

Thus, while the thermodynamic properties of our simulations have been
invalidated, the conclusion that stellar winds are capable of carrying
off accreted angular momentum is not affected.

\section{Conclusion} \label{sec_conclusions}

Based on the emission properties of $\sim 10^6$ K coronal plasmas, we
rule out hot coronal winds as a likely candidate for accretion-powered
stellar winds.  The coronal wind hypothesis fails.  Instead, for mass
loss rates comparable to our fiducial value of $10^{-9} M_\odot$
yr$^{-1}$, the winds must be as cool as $\sim 10^4$ K, where radiative
cooling becomes much less efficient than for a coronal plasma.  At
temperatures this low, the pressure force cannot overcome the gravity
of the star, and accretion-powered winds are thus not driven by
thermal pressure.

To date, possibly the best observational evidence for
accretion-powered stellar winds from CTTSs comes from measurements of
blue-shifted absorption features in the He I emission line at 10830
\AA\ \citep[e.g.,][]{edwardsea03, edwardsea06, dupreeea05}.
Furthermore, radiative transfer modeling by \citet{kurosawa3ea06}
suggests that a stellar wind may contribute significantly to the
H$\alpha$ line profile.  At densities where collisions between
particles are important, both He I and H start to become substantially
ionized at a temperature of a few times $10^4$ K.  If the wind is much
hotter than this, it may be difficult to explain the prominence of He
I and H I features in observed spectra \citep[see
also][]{johnskrullherczeg07}.  Thus these works also support the
conclusion that the winds are much cooler than a coronal plasma.

Accretion-powered stellar winds remain a promising scenario for
solving the stellar angular momentum problem.  But, the question
remains, what is the nature of these winds?  What drives them?
Possible scenarios include Alfv\'en wave driving \citep{decampli81},
episodic magnetospheric inflation \citep{goodson3ea99, matt3ea03}, and
reconnection X-winds \citep{ferreira3ea00, ferreira3ea06}.

\begin{acknowledgments}
We thank the organizers for a stimulating conference and for the
opportunity to present this work.  Gibor Basri deserves credit for
issuing a friendly challenge to our coronal wind hypothesis, six
months prior to this meeting.  He wins the challenge, as it turns out,
since he correctly surmised there would be too much X-ray emission.
Thanks also to J\"urgen Schmitt for making us aware of the CHIANTI
software and database and for discussion about calculating radiative
losses.

CHIANTI is a collaborative project involving the NRL (USA),
RAL (UK), and the following Univerisities: College London (UK),
Cambridge (UK), George Mason (USA), and Florence (Italy).  The
research of SM was supported by the University of Virginia through a
Levinson/VITA Fellowship partially funded by The Frank Levinson Family
Foundation through the Peninsula Community Foundation.  REP is
supported by a grant from NSERC.
\end{acknowledgments}



\end{document}